\documentclass[preprint2]{aastex}
\shorttitle{Broad Emission-Line Response of AGNs}
\shortauthors{Korista \& Goad}
\begin{document}

\title{What the Optical Recombination Lines Can Tell Us About the\\
Broad-Line Regions of Active Galactic Nuclei}

\author{Kirk T.\ Korista}
\affil{Department of Physics, Western Michigan University}
\affil{Kalamazoo, MI 49008-5252}
\email{korista@wmich.edu}

\and

\author{Michael R.\ Goad}
\affil{Department of Physics and Astronomy, University of Southampton}
\affil{Highfield, Southampton, SO17 1BJ, England, UK}
\email{mrg@astro.soton.ac.uk}

\begin{abstract}

We investigate the effect of a global change in the ionizing continuum
level on the behavior of the strong optical broad emission lines seen in
spectra of the nuclear emission-line regions of active galactic nuclei
(AGN), including the Balmer lines, \ion{He}{1} $\lambda$5876, and
\ion{He}{2} $\lambda$4686. Unlike most of the prominent heavy element
lines found in the UV, the optical hydrogen and helium recombination
lines' emissivities are strongly dependent on the incident continuum
flux, since these lines arise out of excited states whose {\em optical
depths depend on the incident flux of photons}. Using photoionization
calculations we determine the luminosity-dependent responsivities,
$\eta(r,L(t)) = \Delta \log L_{line}/\Delta \log L_{cont}$, of these
lines for a general model of the broad emission line region (BLR),
with the purpose of establishing them as important probes of the
physical conditions within the BLR of AGNs. The dependence of these
lines' emissivities on the incident photon flux invokes a dependence
in their responsivities on distance from the central continuum
source. In particular, the responsivities of these lines are generally
anticorrelated with the incident photon flux. Thus, their responsivities
vary with distance within the BLR for a fixed continuum luminosity
and change with time as the continuum source varies. Consequently,
after correcting for light-travel-time effects the response of the
Balmer and optical helium lines should generally be strongest during low
continuum luminosity states. Responsivity that depends on photon flux and
continuum state may explain a number of outstanding problems currently
under investigation in broad-line variability studies of these and
other emission lines. These include the origin of the intrinsic Baldwin
effect, measurements of luminosity-dependent lags (a ``breathing'' BLR)
and luminosity-dependent variations in: integrated broad emission-line
flux ratios (including \ion{He}{2} $\lambda$4686/H$\beta$), broad line
profile shapes, and radial velocity-dependent intensity ratios. The broad
H$\alpha$/H$\beta$ and H$\beta$/\ion{He}{1} flux ratios and the Balmer
emission-line responsivity are observed to decrease from the line center
to the line wings. These, along with our findings, lead to the conclusion
that the BLR velocity field diminishes with increasing distance from the
central continuum source. This is consistent with recent reverberation
studies that find a relationship between the emission-line lag and rms
profile width for multiple lines in individual AGN, which implies that
the velocity field is dominated by a central massive object. Finally,
the responsivity of ionization-bounded clouds can account for much of
the observed behavior of the optical recombination lines (e.g., the weak
response of the Balmer line wings) previously attributed to a substantial
contribution from matter-bounded clouds at small BLR radii.


\end{abstract}

\keywords{galaxies: active---galaxies: nuclei---galaxies: Seyfert---
(galaxies:) quasars: emission lines---line: formation---line: profiles}


\normalsize

\section{INTRODUCTION}

Historically, attempts at interpreting the observed variability behavior
of the broad emission lines in active galactic nuclei (AGN) have tended
to regard the broad emission line region (BLR) as a stationary entity
(non-evolving). By this we mean that any variability in the observed
emission-line intensities has been attributed solely to reverberation
(light-travel-time) effects within a spatially extended BLR. Thus,
variations in the measured BLR sizes and the amplitude of the line
response have been assumed to arise primarily through differences in
the temporal variability of the ionizing continuum, which manifests
as a difference in shape and FWHM of the continuum autocorrelation
function (ACF) from one event to the next (e.g., P\'{e}rez, Robinson, \&
de~la~Fuenta 1992). In essence, higher frequency continuum variations
are better able to probe smaller BLR sizes, while larger BLR sizes are
better probed by lower frequency continuum variations.

While a difference in the continuum variability time scale (e.g.,
the width of the continuum ACF) explains in part some of the observed
variability seen in the broad emission lines in AGNs, it is by no
means the complete picture. Indeed, more sophisticated photoionization
calculations (e.g., see Krolik et~al.\ 1991; O'Brien et~al.\ 1995;
Bottorff et~al.\ 1997; Kaspi \& Netzer 1999; Horne, Korista, \& Goad 2003)
have modeled the line variability by including the local emission-line
response to changes in the overall luminosity and/or shape of the
ionizing continuum.  The models presented by Carroll (1985) and Netzer
(1991) anticipated the non-linear response of some of the emission lines.

In this paper we illustrate, using a general description of the BLR,
how large changes in the mean continuum level may explain much of the
observed variability behavior of the optical recombination lines. In
$\S$~2 we describe the grid of photoionization models used in this
study. In $\S$~3 we describe the physical origin of line responsivity and
its impact on the following issues related to BLR variability: the origin
and slope of the intrinsic Baldwin effect ($\S$~3.2; e.g., Kinney, Rivolo,
\& Koratkar 1990); variations in the measured continuum--emission-line
time-delays ($\S$~3.3; e.g., Peterson et~al.\ 2002), variations in the
emission line flux ratios ($\S$~3.4; e.g., Balmer decrement: Tran,
Osterbrock, \& Martel 1992; H$\beta$/\ion{He}{2}~4686: Peterson \&
Ferland 1986); changes in profile shape ($\S$~3.5; e.g., Wanders \&
Peterson 1996); and velocity-dependent line intensity ratios ($\S$~3.6;
e.g., Stirpe, de~Bruyn, \& van~Groningen 1988). We emphasize that the
variations in these quantities discussed here are {\em not} a consequence
of reverberation effects within a finite-sized BLR, although they are
intimately related. Rather, they are due to global changes in the mean
ionization state of the BLR gas. In $\S$~4 we discuss the impact of these
findings on the purported contribution of a very broad optically thin
component to the emission-line response, and other related issues. It
is our hope that many of these observed emission-line variations and
trends reported in the literature may be examined in a new light. We
summarize our main findings in $\S$~5.

\section{PHOTOIONIZATION MODELS}

\subsection{The Grid of Photoionization Models}

For the purposes of demonstration, we adopted the grid of photoionization
computations presented in Korista \& Goad (2000; hereafter KG00) that
they used to model the strong UV broad emission lines of NGC~5548. The
reader is directed there for details; here we mention some of the salient
features of their grid and adopted model.

Using Ferland's spectral synthesis code, Cloudy (v90.04; Ferland 1997;
Ferland et~al.\ 1998), KG00 generated a grid of 3249 photoionization
models of broad line-emitting entities, here assumed to be simple, single
total hydrogen column density slabs ($N_H = 10^{23}~\rm{cm^{-2}}$),
each of which has constant gas density (for simplicity) and a clear
view to the source of ionizing photons. The grid dimensions spanned 7
orders of magnitude in total hydrogen gas number density, $7 \le \log
n_H (\rm{cm}^{-3}) \le 14$, and hydrogen-ionizing photon flux, $17 \le
\log \Phi_H (\rm{cm^{-2}~s^{-1}}) \le 24$, and stepped in 0.125 decade
intervals in each dimension. We refer to the plane defined by these
two parameters as the density-flux plane. The choice of spectral energy
distribution (SED) is described in KG00 (see also Figure~A1 in Horne,
Korista, \& Goad 2003). In the Appendix we discuss the sensitivity of
the results presented here to these and other assumptions.

In the top four panels of Figure~1 we show the equivalent width (EW)
distributions of the prominent optical broad emission lines of H$\alpha$
$\lambda$6563, H$\beta$ $\lambda$4861, \ion{He}{1} $\lambda$5876, and
\ion{He}{2} $\lambda$4686, namely, those of primary interest in the present
paper. These EWs are all referenced to the same incident continuum flux
at 1215~\AA\, and {\em are a measure of the efficiency by which ionizing
continuum photons are converted into line photons}. The peak in the EW
distribution for each line is denoted by a triangle. Note that the
EWs of the Balmer and \ion{He}{1} lines increase with increasing density,
especially at the lower values of the photon flux, illustrating the
effects of collisional excitation from excited states (Ferland \& Netzer
1979). In every case, the line EWs decline in the direction of increasing
ionizing photon flux for values greater than those marking the peak
EWs: $\log \Phi_H = 17.125$, 18.250, 19.125, and 20.750 for H$\alpha$,
H$\beta$, \ion{He}{1}, and \ion{He}{2}, respectively. The rapid decline in
the line EWs of the two Balmer lines for $\log \Phi_H \approx \log n_H +
10.25$ is due to the hydrogen in the fixed--column density slabs becoming
fully ionized, a subject we discuss later. The two helium lines do
likewise at slightly larger values of $\log (U_Hc) \equiv \log \Phi_H -
\log n_H$. The upper left corners of these diagrams represent gas near the
Compton temperature. Detailed discussions of these and other EW contour
plots in the density-flux plane can be found in Korista et~al.\
(1997). In order to aid in comparison and to serve as a fixed point of
reference, the star symbol in each of the panels shows the location of
the standard BLR parameters from Davidson \& Netzer (1979).

The EW distributions of two other hydrogen recombination lines, H$\gamma$
$\lambda$4340 and Ly$\alpha$ $\lambda$1216\footnote{This is a sum
of Ly$\alpha$ and minor contributions from \ion{O}{5}] $\lambda$1218
and \ion{He}{2} $\lambda$1216.}, are plotted in the bottom panels of
Figure~1. We refer to them from time to time for purposes of comparison.

\subsection{A Simple Model for the BLR of NGC~5548: A Test Case}

To model the mean UV spectrum of NGC~5548 from the 1993 {\em HST}
campaign (Korista et~al.\ 1995; hereafter K95) KG00 took the ``locally
optimally-emitting clouds'' (LOC) approach of Baldwin et~al.\ (1995) by
summing the emission from a weighted distribution of ``clouds'' along
the two dimensions of gas density and radius; KG00 assumed constant
column density clouds of $10^{23}$~cm$^{-2}$. The cloud distribution
functions in gas density and radius are described in KG00. In brief,
the model-integrated spectra included clouds spherically distributed
around the continuum source whose gas densities\footnote {Given the
choice of a single cloud column density and the assumption of constant
hydrogen gas densities within the cloud, the cloud radial thicknesses
range over $10^{11} - 10^{15}$~cm.} and distances from the central source
of ionizing photons spanned $8 \leq \log n_H~(\rm{cm^{-3}}) \leq 12$ and
1--140 lt-days; very highly ionized clouds ($\log U_Hc > 11.25$)
were excluded from the integration (see KG00 for details). Asymmetric
geometric effects, such as emission-line anisotropy (Ferland et~al.\
1992; O'Brien et~al.\ 1994), were not considered. KG00 showed that this
simple model with widely distributed cloud properties was also generally
successful in reproducing the observed UV broad emission-line variability.

With the choice of cosmological parameters and adopted Galactic
reddening along the line of sight to NGC~5548 from KG00, the mean
value of $\lambda\/F_{\lambda}(1350)$ continuum flux from the 1993
{\em HST} campaign (K95) corresponds to $\log \lambda\/L_{\lambda\/1350}
(\rm{ergs~s^{-1}})$ $\approx 43.54$. Using the central continuum SED adopted
by KG00, the corresponding mean hydrogen-ionizing luminosity from the
1993 {\em HST} campaign is $\log L_{ion} \rm{(ergs~s^{-1})} \approx
44.26$. At this luminosity a hydrogen-ionizing photon flux $\log \Phi_H
(\rm{cm^{-2}~s^{-1}}) = 20$ corresponds to a distance from the continuum
source of $R \approx\/ 12.6 (75/H_{\rm{o}})$ lt-days, and the model BLR
spans $\log \Phi_H (\rm{cm^{-2}~s^{-1}}) \approx 17.9$ to $\approx 22.2$
in this mean continuum state. In this paper we consider two historical
extrema in the $\lambda\/F_{\lambda}(1350)$ continuum flux of NGC~5548
about the above mean level. In 1992, the continuum went into a very low
state, $\log \lambda\/L_{\lambda\/1350} (\rm{ergs~s^{-1}}) \approx 42.90$,
while a high state of $\log \lambda\/L_{\lambda\/1350} (\rm{ergs~s^{-1}})
\approx 43.81$ was recorded during the 1989 campaign. This is a factor
of about 8.2 change in the continuum flux at 1350~\AA\/, and we assumed
that the SED did not change shape.

While we do present results based on a simple LOC model of the BLR
in NGC~5548, they should be representative of most general classes of
models describing the broad emission lines. The major trends should be
largely model-independent, while the details should be diagnostic of
the physical conditions within individual BLRs.

\section{EMISSION-LINE RESPONSIVITY}

Summing the line emission along the cloud gas density distribution at each
radius, we can derive an effective cloud surface flux of the emission
lines at that radius (see KG00; Baldwin et al.\ 1995). Figure~2 shows
the surface flux of H$\beta$ as a function of radius from the central
continuum source for each of the two continuum states. The slope of cloud
emission-line surface flux plotted against the ionizing flux ($\propto
L_{ion}/r^2$) is a measure of the effective radial responsivity of the
emitting gas for small amplitude (or equivalently, short time scale)
changes in the ionizing continuum (see Goad et~al.\ 1993). Similarly,
the difference in the two surface flux curves plotted in Figure~2
(in $\log_{10}$ space) relative to the logarithm of the factor of
8.2 in the incident continuum flux measures a ``global'' response to
long-term, large amplitude, changes in the continuum state. We define
this responsivity parameter as \begin{equation} {\eta(r) = \frac{\Delta
\log F(r)_{line}}{\Delta \log \Phi_H}}.\end{equation} This definition
of line responsivity converges to the definition given in Goad et~al.\
(1993) for very small variations in the continuum flux\footnote{In an
alternative approach to analyzing the emission-line response to continuum
variations, Sparke (1993) compared the width of the emission-line ACF
with that of the continuum ACF.}. Note that the upper curve in Figure~2
is essentially the lower curve shifted along the x-axis by $0.5 \times
\log(8.2)$ decades. A logarithmic slope of $-2$ in Figure~2 corresponds
to $\eta(r) = 1$, whereas flatter slopes correspond to $\eta(r) <
1$. When integrated over the full line-emitting region, what we call
the effective responsivity, $\eta_{eff}$, is the logarithmic slope in
the relation $L_{line} \propto L_{cont}^{\eta_{eff}}$.

The solid curves in Figure~3 illustrate the global responsivities as
functions of radius of H$\beta$ ({\em bottom left}), H$\alpha$ ({\em top
left}), \ion{He}{1} $\lambda$5876 ({\em top right}), and \ion{He}{2}
$\lambda$4686 ({\em bottom right}) for this large change in continuum
state. We emphasize that the major changes in responsivity with radius
are not a result of our choice of fixed-column density clouds; this is
discussed below. The dashed and dotted curves represent the responsivities
for small changes in the continuum flux about the high and low continuum
states, respectively, and these are discussed in $\S$~3.1 and $\S$~3.3. It
is clear from the global responsivities that larger relative variations
of these emission lines' surface fluxes due to changes in the continuum
state can be expected to occur at larger radii where the incident
continuum fluxes are smaller.

We can understand the general trends in line responsivity, with a minimum
number of model dependent assumptions, through a study of Figure~1. We
can relate the responsivity to the gradient in the EW contours in Figure~1
as \begin{equation} { \frac{ d \log F_{line} }{ d \log \Phi_H} = \eta =
\frac{ d \log EW }{ d \log \Phi_H } + 1. } \end{equation} The responsivity
is proportional to the efficiency in converting a {\em change} in the
number of ionizing continuum photons into escaping line photons, and so
is in proportion to the change in EW contours shown in Figure~1 relative
to the incident photon flux. For a fixed scale in cloud distance from
the central ionizing source, the EW contours in Figure~1 shift vertically
upward with a decrease in the continuum luminosity, and shift vertically
downward with an increase in the continuum luminosity. Clouds in the
density-flux plane for which the line EW distribution is either flat or
changes little with the incident photon flux have responsivities that
lie near 1 (i.e., roughly 1:1 line:continuum variations) for that line
(see, especially, \ion{He}{2} $\lambda$4686). Those clouds whose line
EWs increase with increasing incident photon flux have responsivities
exceeding 1 for that line (bottoms of the panels for each of these lines
except Ly$\alpha$). Those clouds whose line EWs decrease with increasing
incident photon flux have responsivities less than 1 for that line: the
steeper the gradient, the lower the responsivity. Finally, where the line
EWs decline very rapidly with incident photon flux in the density-flux
plane (near the central diagonal of the figures), the emission line
responsivities become negative.

To illustrate these effects, in Figure~4 we superpose a responsivity
grey-scale on the EW contours for each of the lines in Figure~1. The
grey-scale representations are: $\eta < 0$ ({\em white}), $0 \leq \eta <
0.5$ ({\em light gray}), $0.5 \leq \eta < 1.0$ ({\em medium gray}),
$1.0 \leq \eta < 1.5$ ({\em medium dark gray}), $1.5 \leq \eta < 2$
({\em dark gray}), and $\eta \geq 2$ ({\em black}, appears only in
\ion{He}{2}). Given that we generally measure $0 < \eta_{eff} < 1$ for
the broad emission lines, this diagram shows how the emissivity
and responsivity together constrain the cloud properties, especially
their distribution in radius from the continuum source.

We now briefly describe how the responsivities of the optical
recombination lines arise. At sufficiently high gas densities and
ionizing photon fluxes, each of these four emission lines' EWs diminish
with increasing value of the ionizing photon flux (Fig.~1). This
indicates a lessening efficiency of the conversion of ionizing continuum
photons to escaping line photons for those clouds lying nearer to the
continuum source. This is primarily due to the increasing populations
and thus optical depths in the excited states of hydrogen and helium
with increasing continuum flux (Netzer 1975; Netzer 1978; Ferland \&
Netzer 1979; Ferland, Netzer, \& Shields 1979; Kwan 1984; Rees, Netzer,
\& Ferland 1989; Ferland et~al.\ 1992; Shields \& Ferland 1993), i.e.,
thermalization at large optical depth. At a given continuum flux, the
optical depth of H$\alpha$ is largest, followed by H$\beta$, \ion{He}{1}
$\lambda$5876, and \ion{He}{2} $\lambda$4686 (Rees et~al.\ 1989). Various
background opacities also play a role (Shields \& Ferland 1993), and
photoionization out of excited states of hydrogen becomes effective for
densities exceeding $\sim 10^{10}$~cm$^{-3}$. In fact, those high density
clouds lying above roughly the EW(H$\alpha) = 5$~\AA\/ contour (0.7
dex) in Figure~1 lack a hydrogen ionization front for this reason. As a
consequence of these optical depth effects, the more substantial relative
flux variations in these four emission lines tend to be weighted by clouds
lying at larger radii. This latter point should be true to some extent for
most emission lines, since they too suffer from diminished efficiency in
the conversion of ionizing continuum photons into escaping line photons
at smaller radii where the gas densities of clouds that emit the lines
must be higher. Refer to Figure~2 of KG00 for the EW contours in the
density-flux plane for the stronger UV emission lines. The \ion{Mg}{2}
$\lambda$2800 emission line is most like the Balmer lines in this respect,
and we would expect that line to have a similarly low responsivity (see
Figure~2b in KG00). This is indeed generally observed (e.g., see Clavel
et~al.\ 1991). Clouds fully ionized in hydrogen (such as those lying
near or within the rapidly declining EW contours running diagonally in
the density-flux planes in Figure~1), with their very low ($\sim$~0)
or even negative responsivities, might in principle be arranged in such
a way as to substantially reduce the line responsivity at the larger
radii. However, these low efficiency clouds will not contribute much
light and therefore will add little to the integrated line response
unless their covering fractions are very large.

For time-scales of the order of the light-travel time across the
BLR, the power spectrum of the continuum fluctuations coupled with
light-travel time effects across a finite geometry will also factor into
the variations of an emission line's flux and perhaps profile. However,
{\em it is the responsivity that establishes the outer envelope of these
variations}. Light travel time effects can only act to reduce or dilute
the observed line response. We reiterate that the variations in the line
intensity ratios and line profiles presented here are {\em not} due to
light travel time effects (e.g., see Robinson, P\'{e}rez, \& Binette
1990; P\'{e}rez, Robinson, \& de~la~Fuente 1992; Pogge \& Peterson 1992)
or other geometrical effects (e.g., line anisotropy, O'Brien, Goad, \&
Gondhalekar 1994), but are rather due to differences in the time-averaged
photon flux incident on a spatially extended BLR and their effect on
the line emissivity. Krolik et~al.\ (1991), Pogge \& Peterson (1992),
and Gilbert \& Peterson (2003) have found that reverberation effects
on the $L_{line} \propto L_{cont}^{\eta}$ relation can be largely
removed by correcting the emission-line light-curves for their mean
time-delays. If this is indeed the case, then any residual line --
continuum flux correlations must be primarily due to the responsivity
of the gas in the BLR, as we have defined here. However, if a line's
transfer function has a long tail in time delay (in comparison to the
important continuum variability time scales), simply removing the lag
may not sufficiently correct for this geometric dilution of the line
response. The observed lag-corrected emission-line response will then be
a function of the line's responsivity, as defined here, and the prior
continuum history. More advanced methods, such as that introduced by
Horne et~al.\ (2003), may then be needed to disentangle the effects.

Finally, it is apparent from Figs.~1, 3, and 4 that \ion{He}{2}
$\lambda$4686 should be the most responsive emission line of the four
optical emission lines discussed here, followed by \ion{He}{1}, H$\beta$,
and lastly H$\alpha$. This is observed (e.g., see Wamsteker et~al.\ 1990;
Dietrich et~al.\ 1993; Kollatschny 2003). The model effective global
responsivities of these four emission lines are listed in column~(2)
of Table~1. To illustrate the trends within the Balmer line series,
we also plot the EW distribution of H$\gamma$ $\lambda$4340 within the
density-flux plane in Figure~1, list its effective responsivity in
Table~1, and illustrate its responsivity in the density-flux plane
in Figure~4. Ly$\alpha$ $\lambda$1216 is also a recombination line, but
with different physics, and so we provide the same information for it.

\subsection{The Emission-Line Response Versus the Local Ionizing Photon
Flux and Continuum Luminosity State}

Figure~3 shows that the responsivity of H$\beta$ is expected to be a
strong function of radius, varying from $\sim$~0.2 in the inner BLR to
$\sim$~1 at the outer radius. More generally, this is indicated by the
gradient in the EW contours with respect to the ionizing photon flux in
Figure~1 (see also Eq.~2), and coarsely illustrated in Figure~4. Another
interesting point raised by Figure~3 is that the H$\beta$ line response
should be significantly stronger during low continuum flux states than
during high ones for similar continuum fluctuation amplitudes. This is
illustrated in Figure~3 by the dotted and dashed curves representing the
low and high continuum state H$\beta$ line responsivity, respectively,
computed for small (0.1 dex) variations about the high and low states. The
effective responsivities in the high and low states are given in columns
(3) and (4) of Table~1. K.\ Horne (2002, private communication) and
Cackett \& Horne (2005) recovered the luminosity-dependent delay map
for the 13 yr monitoring campaign of the H$\beta$ broad emission line
of NGC~5548, and found a similar relation between the amplitude of the
line response with continuum state.


Figure~3 and columns (3) and (4) of Table~1 shows that all of these
lines display generally larger effective responsivities at larger
distances (lower incident photon fluxes) and during lower continuum
states. Additionally, the amplitude of the change in effective
responsivity with continuum state differs substantially between the Balmer
and helium lines, and this should be diagnostic of the physical conditions
within the BLR. By contrast, Ly$\alpha$ shows very little variation in
responsivity with cloud parameters (Fig.~4), and as a consequence shows
a very modest change in $\eta_{eff}$ with continuum state (Table~1).

We elucidate the change in the radial responsivity with continuum state
shown in Figure~3, using \ion{He}{2} $\lambda$4686 by way of example. In
both continuum states, the responsivity at large radii is nearly
independent of radius and continuum state. This is simply a reflection of
the general lack of dependence of the line EW with incident continuum flux
for $\Phi_H \lesssim 20.5$ (see Figure~4). This is in contrast to the case of
the Balmer lines, especially H$\alpha$. Note also in Figure~4 that most of
the variations of $\eta$(\ion{He}{2}) with incident photon flux occur for
$\Phi_H \gtrsim 20.5$; here in particular $\eta$ declines with increasing
$\Phi_H$. This dependence of the responsivity on incident continuum flux
makes this line's radial responsivity at small distances quite dependent
on the continuum state as well. These effects are also illustrated in
Figure~3. The ``knee'' in the radial responsivity of \ion{He}{2} occurs
at the radius corresponding to $\Phi_H \approx 20.5$ for each continuum
state, or $\log r \approx 16.25$ (7 lt-days) in the mean continuum
state. Simply put, strong anti-correlations between the line EW and the
incident photon flux will correspond to anti-correlations between $\eta$
and the incident photon flux, which in turn will result in similar
anti-correlations between $\eta$ and the continuum state. Likewise,
$\eta_{eff}$ will anti-correlate with the continuum state, as weighted by
the line's radial luminosity function. Based on single power-law cloud
distribution models, O'Brien, Goad, \& Gondhalekar (1995) also found
``a decrease in the mean response when the continuum flux increases.''

By definition, a time-varying responsivity $\eta (r,L_{cont}(t))$ implies
that the line transfer function, $\Psi(\tau,v$), is time-dependent (i.e.,
non-stationary). In this case the time dependence is due to changes in
the time-averaged photon flux incident on a spatially extended BLR. We
therefore expect the measured lags of the emission lines and their profile
shapes to depend on the luminosity state of the continuum ($\S$~3.3
\& $\S$~3.4). We note that unless the velocity field is self-similar
in radius (Goad et~al.\ 1999), a profile shape that depends on the
continuum luminosity state fails a hysteresis test (see Wanders 1994;
Perry, van~Groningen, \& Wanders 1994). This is true without invoking a
redistribution of material within the BLR or other geometrical effects
as suggested by Wanders \& Peterson (1996).

\subsection{The Intrinsic Baldwin Effect}

Detailed analyses of the AGN broad emission-line response to continuum
variations in spectrophotometric monitoring campaigns show that in
all cases the responsivity in most measured lines is $< 1$, even
after accounting for light travel-time effects (e.g., Krolik et~al.\
1991). Referring to Eq.~2, we see that this requires the line equivalent
width to diminish with increasing continuum flux and vice versa:
the intrinsic Baldwin effect ($EW_{line} \propto L_{cont}^{\beta}$;
see Kinney, Rivolo, \& Koratkar 1990; Krolik et~al.\ 1991; Pogge \&
Peterson 1992; Peterson 1997; Gilbert \& Peterson 2003). In concurrence
with Gilbert \& Peterson the present work suggests that the origin of
the intrinsic Baldwin effect is due either in whole or in large part
to the line's responsivity being less than 1 throughout the majority
of a distribution of largely ionization-bounded clouds (as shown in
Figure~1), i.e., without the {\em necessity} of a large population of
matter-bounded clouds (Shields, Ferland, \& Peterson 1995), although we
do not rule these out. As discussed above we expect $\eta_{eff} < 1$ for
most lines, and thus most lines should demonstrate an intrinsic Baldwin
Effect, since the intrinsic Baldwin effect index $\beta$ is related to
the responsivity as $\beta = \eta_{eff} - 1$. Based on single power
law cloud distribution models, O'Brien, Goad, \& Gondhalekar (1995)
showed that an intrinsic Baldwin effect can be induced in any line by
changing its responsivity with continuum level. While we concur with
these findings, we find a more general and significant origin for this
effect, namely $\eta_{eff} < 1$ as defined and described in $\S$~3.

Our simple model for the BLR of NGC~5548 predicts an $\eta_{eff}$ for
H$\beta$ of 0.54 in the 1989 high state and 0.77 in the 1992 low state
(columns (2) -- (4), Table~1). That is, we expect $\eta_{eff}$ and so
the slope in the intrinsic Baldwin effect to be a function of continuum
state. Compare the dotted and dashed lines in Figure~3. This predicted range
in $\eta_{eff}$ approximately brackets that measured in the same object by
Gilbert \& Peterson (2003) of $0.53 - 0.65$.\footnote{The logarithmic slope
0.65 was obtained when the value of the background galactic continuum
at 5100~\AA\/ was fixed to the observed value (Romanishin et~al.\ 1995),
while 0.53 was obtained when this background light was allowed to vary in
the fit.} A line's equivalent width is usually measured with respect
to the local continuum, and Gilbert \& Peterson (2003) found that
the underlying continuum becomes bluer when brighter in this object,
such that $L_{5100} \propto L_{1350}^{0.67}$. They then suggested that
since the $\lambda$1350 continuum lies closer to the ionizing continuum,
it might be better to reference the H$\beta$ flux to it, and found the
relation $L(H\beta) \propto L_{1350}^{(0.35 - 0.43)}$. Our simulations
assumed that the SED remains constant as its amplitude changes. Thus,
while it is difficult to make detailed comparisons, we expect that SED
changes are a secondary effect on the line responsivity; we discuss
the impact of a variable SED on line responsivity in $\S$~4.2. The
point remains that line responsivity, as defined here, is likely to be
responsible for much of the observed intrinsic Baldwin Effect.

What can the intrinsic Baldwin effect tell us about the BLR? Because
the EW contours of the Balmer lines in particular in Figure~1 are so
nearly proportional to the incident continuum flux, we suggest that
measurements of the responsivities of these emission lines will point to
the distribution of the optical recombination line-emitting entities
in the density-flux plane. Low density gas ($n_H < 10^{10}$ cm$^{-3}$)
illuminated by low values of the ionizing photon flux ($\Phi_H < 10^{18}$
cm$^{-2}$s$^{-1}$), and gas at any density illuminated by even lower
values of $\Phi_H$ ($< 10^{17.5}$) will emit these optical recombination
lines, but they {\em will not respond appropriately} to variations in the
ionizing continuum. The responsivities of these clouds in these lines
are $\sim$~1 and $> 1$, respectively. Such models would fall far short
of reproducing the observed variations in H$\beta$, {\em independent}
of any considerations of the line luminosity or lag. Next, the observed
mean responsivity of the optical \ion{He}{2} $\lambda$4686 line, $\eta
\sim 0.9$ (see, for example, Dietrich et~al.\ 1993; also Krolik et~al.\
1991), necessarily implies that a significant fraction of this line forms
in gas exposed to high incident ionizing continuum fluxes on average
($\log\Phi_{H} \gtrsim 20.5$~cm$^{-2}$~s$^{-1}$, or $R \lesssim 7$ light
days in NGC~5548), as well as relatively high densities\footnote{Higher
column density clouds would allow for somewhat lower gas densities under
this condition, and vice-versa (e.g., see Korista et~al.\ 1997).}. This
is just as its measured lag indicates (Clavel et~al.\ 1991; K95). Thus,
Figures~3 and 4 predict a positive correlation between the lag of each
of these lines and their responsivity, $\eta_{eff}$, for a particular
continuum state (see also Carroll 1985). So together the luminosity,
lag, {\em and} responsivity (intrinsic Baldwin Effect) of each of the
broad emission lines for a range of continuum states should constrain
the distribution of BLR cloud properties in gas density, distance from
the ionizing continuum source, and column density. The quasar tomography
method of Horne et~al.\ (2003) is particularly well suited to this task,
and we discuss this further in $\S$~4.3.

%

\subsection{The ``Breathing'' Broad Line Region}

Netzer \& Maoz (1990) reported that different lags for the broad emission
lines of Ly$\alpha$ and \ion{C}{4} were found for each of the three
continuum variability events during the 1989 campaign of NGC~5548 (Clavel
et~al.\ 1991). Based on data from an earlier optical campaign (Netzer
et~al.\ 1990), Netzer \& Maoz also reported a significantly different
lag for H$\beta$ than did Peterson et~al.\ (1991). While acknowledging
that a variable continuum ACF will change the lag, Netzer \& Maoz (1990)
suggested that at least part of the origin of these time-dependent lags
lay in a non-linear emission-line response. Over the full 13 yr monitoring
campaign of NGC~5548 Peterson et~al.\ (2002; see also Peterson et~al.\
1999) found that the H$\beta$ emission-line lag scales with the mean
continuum level, such that $\tau\/ \propto\/ F_{5100}^{0.95}$, albeit
with much scatter. This positively correlated behavior of the lag with
continuum state has been called ``breathing.'' After positing that the UV
continuum is more intimately connected to the driving ionizing continuum
and accounting for the relation that they found between the $\lambda$5100
and $\lambda$1350 continuum bands ($F_{5100} \propto F_{1350}^{0.56}$),
the relation $\tau\/ \propto\/ F_{UV}^{0.53}$ resulted. Peterson et~al.\
(2002) then proposed that $\tau\/ \propto\/ F_{cont}^{0.5}$ might be
a natural consequence of a simple ionization parameter-type argument
that $r \propto L_{ion}^{0.5}$. Cackett \& Horne (2005), too, have found
evidence of breathing in their transfer function analysis of the 13 yr
spectroscopic dataset of Peterson et~al.\ (2002). A breathing BLR has also
been proposed by Goad et~al.\ (1999) to explain line profile variation
differences between \ion{C}{4} and \ion{Mg}{2} in NGC~3516. In this
subsection, we address a likely origin of breathing BLRs and comment on
the conclusions drawn by Peterson et~al.\ (2002) in regards to this issue.

The single power law cloud distribution models of O'Brien, Goad, \&
Gondhalekar (1995) predicted a relation between line lag and incident
continuum level (for a fixed SED) due to a non-linear emission-line
response. The results presented here for the optical recombination lines
based on more general models arrive at the same conclusion. The shift
in the lag as a function of the continuum state is a reflection of an
outward shift in the line's luminosity-weighted radius to gas at larger
radii that is better able to reprocess the increased continuum luminosity
into line emission, and vice versa in low continuum states. The EW
contours in Figs.~1 and 4 shift vertically as the source luminosity
changes, causing corresponding changes in the radial responsivity and
hence responsivity-weighted radii. We therefore expect that elevated
continuum states will be accompanied by an increase in the line lag and
a decrease in the line responsivity.

Breathing will not occur for lines for which the responsivity as
a function of radius and ionizing continuum level is constant. A
constant $\eta(r,L(t))$ implies that the emissivity-weighted and
responsivity-weighted radii are the same (see Goad et~al.\ 1993),
and therefore the measured lag is independent of the continuum
state. Therefore, if after correcting for reverberation effects,
the measured lag varies with luminosity, then so must its effective
responsivity. Although Gilbert \& Peterson (2003) did not note any
evidence for changes in $\eta_{eff}$ with continuum level, B.\ Peterson
(2003, private communication) finds that a detailed analysis of the 13 yr
campaign on a year-to-year basis indicates changes in the line--continuum
flux slope. It is also significant that while BLRs whose lines breathe
must show an intrinsic Baldwin effect for those lines, the converse
is not true. For example, for the case of $\eta = 0.5$ everywhere,
no breathing will occur, but an intrinsic Baldwin effect will be present.

The two triangles in each panel of Figure~3 mark the positions of the
responsivity-weighted radii ($R_{\eta}$) of H$\beta$ and the other three
emission lines for the two continuum states. In the case of H$\beta$, it
is larger in the high continuum state by a factor of $\sim$~1.6 relative
to the low state. This is about the same ratio of luminosity dependent
lags observed in this line in NGC~5548 for the same two continuum states
($\sim$~1.7; Peterson et~al.\ 2002; see also Wanders \& Peterson 1996;
Cackett \& Horne 2005), although we note that the measured lag
is expected to be smaller than $R_{\eta}/c$ by a factor of a few (e.g.,
see P\'{e}rez et~al.\ 1992).

The effective line responsivities for the high and low continuum state
are given in columns (3) and (4) of Table~1. As discussed in $\S$3.1,
the low-state responsivity is higher in every case, and Figure~3
shows that the line's radial responsivity and $R_{\eta}$ shift to
larger distances in higher continuum states. For this particular
model, the responsivity-weighted radius scales as $R_{\eta} \propto
L_{cont}^{\gamma}$, with $\gamma =$ 0.15, 0.23, 0.38, and 0.44 for
H$\alpha$, H$\beta$, \ion{He}{1}, and \ion{He}{2}, respectively. These
relations should be diagnostic of the physical conditions within the
BLR. The general relation proposed by Peterson et~al.\ (2002), above
($\gamma = 0.5$), is expected to hold only if the line responsivity is
constant over a large portion of the BLR with a steep break to lower
responsivity (see for example, \ion{He}{2} in Figure~3), and so is not
the general case. The breathing amplitude (e.g., the value of $\gamma$)
is expected to depend on the responsivity and the extent to which
the responsivity changes with both radius and continuum state in the
vicinity of $R_{\eta}$ (compare dotted and dashed curves for each of
the lines in Figure~3). Using the relation between the $\lambda$5100
and $\lambda$1350 continuum fluxes for NGC~5548 (Gilbert \& Peterson
2003; but see also Peterson et~al.\ for a slightly different value), and
assuming that $L_{1350}$ is a proxy for the driving continuum, K.\ Horne
(2002, private communication) and Cackett \& Horne (2005) find a similarly
weaker relation between the H$\beta$ lag and continuum luminosity state
for the 13 yr campaign of NGC~5548 as do we. In addition, since $\gamma$
is related to the responsivity, it too changes with continuum level
(see also O'Brien et~al.\ 1995).

Finally, we note that the expected larger line response during low
continuum states ($\S$~3.1) will be enhanced by the fact that an
effectively smaller line-emitting region, due to this breathing,
respond more coherently to a given continuum variation. The opposite
hold for variations about high continuum states. The combined effect
can be investigated in data sets such as the 13 yr H$\beta$ light curve
presented in Peterson et~al.\ (2002; their Figure~1): more coherent
and larger amplitude responses in H$\beta$ should generally be found at
low continuum states, while just the opposite should result during high
continuum states. One might also find differences in line responsivity
between two comparably low continuum states that immediately straddle a
high continuum state in time, due to light travel time effects that are
not removed from the second low state by simply correcting for the lag.

\subsection{Integrated Line Flux Ratios vs.\ Continuum State}

In this section, we investigate the expected changes in the integrated
line flux ratios due to changes in the continuum state. It should
now be apparent that such line ratio variations will result simply
because the responsivities differ between the emission lines. Lines
with greater responsivities vary with greater amplitudes with changing
continuum state, and so line flux ratios can be expected to change as
well. Other factors may play a role, but responsivity effects will be
present regardless.

Figure~5 illustrates the H$\alpha$/H$\beta$ line flux ratio varying
across the density-flux plane, from roughly 17 (near the triangle at the
bottom) to approximately 1 (near coordinates: [13.5, 22.5]). The classical
nebular theory of hydrogen Balmer lines (Menzel case~B ) does not apply
to the gas occupying the BLR. For example, at coordinate [12.0, 19.5]
in the density-flux plane (corresponding to a distance of 22 lt-days
in the mean continuum state), case~B predicts H$\alpha$ and H$\beta$
fluxes that are 20.8$\times$ and 30.4$\times$ larger than that emitted
by the cloud. The case~B H$\alpha$/H$\beta$ flux ratio is 2.35, whereas
the ratio predicted by Cloudy is 3.44. Over a significant portion
of the density-flux plane {\em this ratio is inversely proportional
to the incident continuum flux}, as expected from the optical depth
effects mentioned earlier, and in agreement with the findings of Rees,
Netzer, \& Ferland (1989; see also Netzer 1975). This being the case,
it is easy to imagine this ratio changing in response to variations in
the incident continuum luminosity. Thus, differences in responsivity
among the Balmer lines, as illustrated in Figs.~1, 3, 4 and as listed
in Table~1, probably account for most of the observed steepening in the
broad-line Balmer decrement as the continuum drops into low states,
and flattening in high states. There have been many such broad line
Balmer decrement/continuum state correlations reported in the spectra
of variable Seyfert~1 galaxies (see, for example, Antonucci \& Cohen
1983; Ferland, Korista, \& Peterson 1990; Wamsteker et~al.\ 1990; Tran,
Osterbrock, \& Martel 1992). Integrated over this model's full broad
line-emitting region, the H$\alpha$/H$\beta$ flux ratio changes from
4.9 in the low state to 3.7 in the high state. While the particulars
are model dependent, the trend generally is not. The Balmer decrement's
variations with the incident continuum level should be sensitive to the
distribution of clouds in the density-flux plane. Furthermore, we suggest
that a Seyfert~1 galaxy's sub-classification as Seyfert~1.0, 1.5, 1.8,
or 1.9 (as defined in Osterbrock 1989, and references therein) may at
least in part be due to the relative flux of ionizing photons present
within their broad-line regions at the time of observation. That is, any
given spectrum of a Seyfert Type~1 AGN may pass through some of these
sub-classifications as its continuum luminosity fluctuates from high
(Seyfert~1 or 1.5) to low continuum states (Seyfert~1.8 or 1.9). In
high continuum luminosity states, a larger broad-line contribution to
the total (broad + narrow) Balmer line profile will be present and the
broad Balmer line decrement will be flatter. In low continuum luminosity
states, the broad-line contribution to the total Balmer line profile will
be relatively much smaller, and the broad Balmer line decrement will be
steeper. NGC~5548 is generally classified as Seyfert~1.5. However, the
order-of-magnitude continuum fluctuations that occurred over the 13 yr
optical monitoring campaign (Peterson et~al.\ 2002) probably drove the
Balmer emission-line spectrum from Seyfert type 1.8 in 1992 to perhaps
Seyfert type 1.0 in 1998.

As described above, the Balmer lines should show less flux variation with
continuum state than \ion{He}{1}, which in turn should show less variation
than \ion{He}{2}. This trend is generally observed (Peterson \& Ferland
1986; Ferland, Korista, \& Peterson 1990; Wamsteker et~al.\ 1990; Dietrich
et~al.\ 1993; Peterson et~al.\ 1993; Kollatschny 2003). Particularly
interesting is that observed changes in the H$\beta$/\ion{He}{2}
flux ratio may, at least in part, due to {\em differences in the
way these two lines respond to the level of the ionizing continuum flux},
rather than to changes in the shape of the ionizing SED. For the model
presented here, this ratio drops by nearly a factor of 2 from the
low to high continuum states. To a lesser extent, this is likewise true
for the \ion{He}{2}/\ion{He}{1} line ratio, which changes by a factor of
1.3. Such a responsivity effect may have played a role in the observation
of a large increase in the \ion{He}{2} $\lambda$4686 equivalent width in
comparison to H$\beta$ during a large increase in the continuum luminosity
in NGC~5548 (Peterson \& Ferland 1986) and likewise in a similar event
reported by Peterson et~al.\ (2000) in NGC~4051. While the broad line
\ion{He}{2}/H$\beta$ flux ratio is sensitive to the ionizing SED, the
usual photon-counting (Zanstra) argument used to measure the 13.6 to
54.4~eV continuum SED applies strictly when these lines are emitted in
gas whose properties allow for responsivities $\eta_{eff} \approx 1$, as
would be true if case~B applied for both lines. However, these conditions
{\em do not} apply for the Balmer lines emitted within the BLR, but they can
be found within the narrow line regions of AGNs (see the EW distributions
of the recombination lines in Ferguson et~al.\ 1997) or within galactic
nebulae. On the other hand, the \ion{He}{2} emissivity does not stray
too far from that expected from case~B for most conditions expected in
the BLR, as indicated by its comparatively featureless EW distribution
in the density-flux plane for $\log \Phi_H \lesssim 21$ (see also
Bottorff et~al.\ 2002).

Finally, as a matter of general interest we note that the
Ly$\alpha$/H$\beta$ flux ratio should correlate with the continuum
flux, and apparently this is the case in the well-studied AGN
NGC~5548 (Wamsteker et~al.\ 1990). The EW contours of Ly$\alpha$ in
the density-flux plane (see Figure~1, {\em bottom right}) do not
show as strong continuum flux dependencies as do the Balmer lines, and
so $\eta$(Ly$\alpha) > \eta$(H$\beta$) can be expected, with larger
differences to come from gas nearer to the continuum source where
the responsivity of H$\beta$ is small (see also Figure~4). The combined
conclusions of Pogge \& Peterson (1992) and Gilbert \& Peterson (2003)
would indicate an observed relation of $L($Ly$\alpha)/L($H$\beta)
\propto L_{1350}^{0.12-0.24}$ in NGC~5548. Our simple model (see
Table~1) predicts a similar relation $L($Ly$\alpha)/L($H$\beta) \propto
L_{1350}^{0.10}$, which is steeper in high continuum states and shallower
in low ones. Shields \& Ferland (1993) discussed some of the mechanisms
behind the destruction of Ly$\alpha$ emitted by broad-line clouds. Pogge
\& Peterson also found that in Fairall~9 $\eta_{eff}$(Ly$\alpha) \approx
0.71$ for a factor of $\sim$10 change in the observed $\lambda$1338
continuum, similar to the value reported here (0.74, Table~1) for
our simple model. Figure~4 shows that $\eta$(Ly$\alpha$) is nearly
independent of gas density and incident photon flux over a wide range
in these parameters.

\subsection{Line Profiles and Their Variations with Continuum State}

While the dynamics of the gas emitting the broad emission lines are
not yet understood, the importance of a virial component seems to be
reasonably well established (Peterson \& Wandel 2000). Here, we have
adopted the simple relationship \begin{equation} { v(r) = (GM/r)^{1/2},
} \end{equation} assuming randomly inclined circular orbits and a central
mass of $8.3 \times 10^7$ solar masses, chosen to approximately reproduce
the observed broad-line FWHM of H$\beta$ in NGC~5548. This choice of
dynamical model is motivated by simplicity for purposes of demonstration.

The low-state continuum emission-line profiles of the four lines under
study, as predicted by our simple LOC model line emissivity and velocity
field, are shown in Figure~6, normalized to their peak intensities. We note
that the flat tops to these profiles are artifacts of the assumption of
circular orbits, and their extents are determined by the local Keplerian
velocity at the distance of the model's outer radius (140 lt-days) given
the assumed central mass. The EW distributions shown in Figure~1, and most
notably the continuum flux at the onset of significant thermalization,
are roughly reflected in the widths of the emission-line profiles
in Figure~6, as well as in their observed lags (Dietrich et~al.\ 1993;
Peterson \& Wandel 1999, 2000; Fromerth \& Melia 2000). \ion{He}{2}
is the broadest line (with the shortest lag), followed by \ion{He}{1},
then H$\beta$, and finally H$\alpha$ (with the longest lag). This is as
generally observed in broad-line AGNs (Stirpe 1990, 1991; Crenshaw 1986,
Shuder 1982, 1984; Osterbrock \& Shuder 1982; Boroson \& Green 1992;
Kollatschny 2003). Krolik et~al.\ (1991), Peterson \& Wandel (1999,
2000), and Onken \& Peterson (2002) present highly correlated lag --
line width relations that indicate virial-like motions in the vicinity
of a supermassive central object. The FWHM of these four emission lines,
as well as H$\gamma$ and Ly$\alpha$, are listed in columns~(5) and (6)
of Table~1 for the high and low continuum states, respectively. The
trend in line widths is in general agreement with those found in the
models presented by Rees, Netzer, \& Ferland (1989).

Since the lines under study here are recombination lines involving just
one or two electrons, they are emissive over very broad areas within the
density-flux plane, unlike the collisionally excited metal lines
which are very sensitive to the ionization parameter. The Balmer
and \ion{He}{1} lines span essentially equal areas of significant
emissivity within the density-flux plane {\em without regard to
ionization parameter}, and that of \ion{He}{2} does likewise for $\log
U_H \gtrsim -3.5$ (see Figure~1). While it is also true that a cloud of a
given column density will become fully He$^{++}$ at a higher ionization
parameter than for H$^+$, this difference is small (see Figure~1). By far
most of the differences in emissivity among the Balmer and \ion{He}{1}
lines are due to the optical depth effects discussed above, and this
is also true for the \ion{He}{2} lines except for clouds with $\log U_H
\lesssim -3.5$. Stated another way, if all of these lines were emitted with
their case~B emissivities within the BLR, much smaller differences in
their profiles and lags would be expected.

Figure~7 shows the long term (global) responsivities for each of the
four emission lines as a function of radial velocity across their
line profiles, for our simple velocity field. Because of the larger
response in the line cores, the broad emission-line component of the
Balmer lines should become narrower/peakier in higher continuum states,
and broader/less peaky during low continuum states (cf., H$\alpha$
variations in NGC~5548; Stirpe et~al.\ 1988); refer also to the model
high and low continuum state FWHMs as listed in columns (5) and (6)
in Table~1. In fact for this particular model, a continuum state that
is a factor of 12$\times$ brighter than the low state\footnote{Such a
high state in NGC~5548 was recorded in 1984 (Wamsteker et~al.\ 1990)
and again in 1998 (Peterson et~al.\ 2002).} produces a FWHM in H$\beta$
that is 1000 $\rm{km s^{-1}}$ narrower than that in the low state. The
response across the profile of \ion{He}{2} is {\em much higher}, although
again this emission line would be expected to be somewhat narrower during
higher continuum states. The response across the profile of \ion{He}{1}
is not as broad as \ion{He}{2} and not as peaky as the Balmer lines. In
every case, {\em we expect the response in the line cores to be stronger
than in the line wings}. The relative change in the FWHM of the emission
line with continuum state is effectively a reflection of the ``breathing''
BLR as caused by changes in the line responsivity, as we discussed in
$\S$~3.3. O'Brien, Goad, \& Gondhalekar (1995) similarly linked non-linear
emission-line response to line profile variability, although this general
conclusion was based on single power law cloud models.



\subsection{Radial Velocity-Dependent Emission-Line Ratios and Their
Variations with Continuum State}

Since the responsivity of these lines is expected to be radially dependent
and this function differs between emission-line species, we expect radial
velocity-dependent variations in the line intensity ratios. These effects
are illustrated in Figure~8, which shows the velocity-dependent Balmer
decrement (H$\alpha$/H$\beta$) for the low-state ({\em dotted line})
and high-state ({\em solid line}) continuum. Note that in both cases the
Balmer decrement is steeper in the line core than in the line wings,
and the Balmer decrement shows larger changes with continuum level in
the line core (from 5.6 to 4.0), than in the line wings (from 3.0 to
2.7). The first of these effects has been well-established (see, for
example, Shuder 1982, 1984; Osterbrock \& Shuder 1982; van~Groningen
1984, 1987; Crenshaw 1986; Stirpe 1990, 1991; Kollatschny 2003), and
we are aware of just one study of the time-dependent variability of the
Balmer decrement as a function of radial velocity (Stirpe, de~Bruyn, \&
van~Groningen 1988). Both behaviors imply that {\em the velocity field
must decrease with increasing radius from the central source}. As just
presented above, a weaker response in an individual line's wings in
comparison to its core (Fig.~7) would also argue for such a velocity
field. These findings are consistent with recent reverberation studies
that find a relationship between the emission-line lag and rms profile
width for multiple lines in individual AGNs, which implies that the
velocity field is dominated by the central massive object (Peterson \&
Wandel 1999, 2000; Fromerth \& Melia 2000).

The results presented here also predict a smaller H$\beta$/\ion{He}{1}
intensity ratio in the line wings than in the core (Fig.~8). The EW
contours in Figure~1 coupled with a velocity field that diminishes with
increasing distance from the central source provide possible explanations
for this generally observed phenomenon (Shuder 1982, 1984; Osterbrock \&
Shuder 1982; Crenshaw 1986; van~Groningen 1984, 1987; Rees, Netzer, \&
Ferland 1989; Kollatschny 2003). These lines' EWs peak up at similarly
high densities, but the \ion{He}{1} line peaks up at significantly
higher photon fluxes (smaller radii), because of its lower optical
depth. This is responsible for this line's relatively stronger wings
(see also Netzer 1978; Kwan 1984; Carroll 1985; Carroll \& Kwan 1985).



Whether or not the specific dependencies of the
continuum state/responsivity effects described in this section are
present in the spectra of NGC~5548 or other variable Seyfert~1 AGNs is not
particularly important. We discuss the sensitivity of line responsivity
to model assumptions in the Appendix. Instead, the emphasis of this paper
lies mainly in illuminating the mostly untapped wealth of the potentially
powerful physical diagnostics present in high quality spectrophotometric
data sets of the optical recombination lines, and that their emissivities'
sensitivity to the continuum flux makes them particularly effective
diagnostics of the radially dependent physical conditions within the BLR.

\section{DISCUSSION}

\subsection{Optically Thin Gas in the So-Called Very Broad Line Region}

That the BLR may contain a significant component of ionized gas that is
optically thin to hydrogen-ionizing photons has been suggested by investigators
over many years. In nearly every case there is an association of this
gas with that emitting the very broad wings of the lines. Most recently,
Sulentic et~al.\ (2000) reported that in the quasar PG~1416$-$129 the
FWHM of the broad emission-line profile of H$\beta$ increased by 50\%
from a high to a low continuum state separated in time by 10 yr. Their
interpretation of this is that the wings of the Balmer lines are dominated
by emission from an optically thin ``very broad line region'' (VBLR),
while the core of the line is dominated by more highly variable optically
thick gas. These are usually thought of as two distinct line-emitting
regions. Ferland, Korista, \& Peterson (1990), Peterson et~al.\ (1993),
and Corbin \& Smith (2000) made similar arguments based on Balmer
line -- continuum variations, while Gondhalekar (1987, 1990), O'Brien,
Zheng, \& Wilson (1989), \& P\'{e}rez, Penston, \& Moles (1989) found
less variable wings in Ly$\alpha$ and \ion{C}{4} $\lambda$1549 of higher
redshift quasars. Kassebaum et~al.\ (1997) found similarly stronger
variations in the line core than wings of H$\beta$ in Mrk~335. It
should be noted that while the conclusions regarding the core versus wing
variations from monitoring campaign studies such as Peterson et~al.\
(1993) and Kassebaum et~al.\ (1997) were based on analyses of {\em
time-resolved} spectral observations of the continuum and H$\beta$, most
of the above studies relied on very few, often just two, observation
epochs. Without sufficient time resolution of the variability, it is
difficult to assess the actual line response. Additional arguments
for an optically thin line-emitting region have been put forth through
the analyses of line flux ratios in the profile wings: optical Balmer
and \ion{He}{2} lines (Marziani \& Sulentic 1993; Corbin 1997), Balmer
and \ion{C}{4} $\lambda$1549 (Corbin 1995), Balmer and Ly$\alpha$ line
profiles (Zheng 1992), and Balmer and \ion{O}{1} $\lambda$8446 profiles
(Morris \& Ward 1989).

While at any given continuum state the broad emission-line region may well
contain clouds fully ionized in hydrogen, as the present model does, the
results presented here suggest that there is no great need for a major
line-emitting component consisting of vast amounts of optically thin
gas. Most of the effects mentioned just above regarding the Balmer and
helium recombination lines can be explained by the effects we described
in $\S$~3: weak Balmer line emission and response need not originate in
clouds optically thin to the Lyman continuum. While an emission line
such as \ion{Ne}{8} $\lambda$774 is likely to be emitted mainly within
clouds that lack a hydrogen ionization front (e.g., see Hamann et~al.\
1998; Korista et~al.\ 1997), we have also pointed out that by their
very nature fully-ionized clouds are inefficient emitters of hydrogen
recombination lines (refer again to Figure~1), and so are unlikely to be
important energetically even if their covering fraction is large. This
is especially true in the face of a variable ionizing continuum source;
clouds will more likely be either ionization-bounded or too overionized
to emit Balmer lines of any significance. It is also significant that
the matter-bounded clouds in Figure~4 either have small positive ({\em
lightest gray})\footnote{The {\em only} clouds in Figure~4 whose hydrogen
line responsivities are $\approx$~0 lie above the diagonal line $\log
(U_Hc) \approx 10.25$ and also within the lightest gray shaded region.}
or negative (white) responsivities. Even if present, the latter type
clouds never make their presence known in these emission lines, which
is not surprising given their extremely low emitting efficiencies.

In regards to the comparisons of Balmer and Ly$\alpha$ line profiles, as
mentioned in $\S$~3.4, the EW contours of Ly$\alpha$ in the density-flux
plane (see Figure~1, {\em bottom right}) do not show the strong continuum
flux dependencies of the Balmer lines. We suspect that this, along with
the presence of a radially decreasing velocity field, accounts for the
very large Ly$\alpha$/H$\beta$ ratios found in the emission-line wings by
Zheng (1992), rather than these ratios being due to a major constituent
of optically thin emitting gas.

Finally, we also point out that many of the above studies assumed that
case~B emissivities apply to the hydrogen lines emitted by optically thick
clouds in the BLR and attributed significant deviations from expected
ratios to an optically thin emitting region. We conclude that a separate
optically thin VBLR may be unnecessary, and in any case for the Balmer
lines is probably energetically unimportant. The wings of the hydrogen
recombination lines likely represent nothing more than the inner BLR
where the Balmer line emission is inefficient and the responsivity low.


\subsection{Effects of an SED that Varies with Continuum Level}

Paltani \& Courvoisier (1994) and Paltani \& Walter (1996) showed
unambiguously that the UV continuum becomes bluer when the flux is higher
in variable AGNs. This effect has also been found in the Sloan Digital
Sky Survey quasar sample (Vanden Berk et~al.\ 2004). In the particular
case of NGC~5548, Maoz et~al.\ (1993) reported that, after accounting for
the Balmer continuum and UV \ion{Fe}{2} emission, the 1800--2400~\AA\/
continuum becomes bluer with increased continuum flux. Additionally,
Peterson et~al.\ (2002) and Gilbert \& Peterson (2003) found that
after careful removal of the contaminating non-variable stellar optical
continuum flux, the UV-optical continuum in NGC~5548 becomes bluer as
it becomes brighter: $F_{5100} \propto F_{1350}^{0.67}$. Extrapolation
of this relation to higher energies might indicate a hardening of the
incident ionizing continuum during brighter continuum states. Marshall
et~al.\ (1997) found larger amplitude variations in the extreme
ultraviolet ($\sim$~100~\AA\/) than at 1350~\AA\/ during a week-long
period of the 1993 NGC~5548 monitoring campaign. Alternatively, as
suggested by Paltani \& Walter (1996), it may indicate the presence
of a constant (or more slowly varying) continuum or pseudocontinuum
component whose contribution is greater at longer wavelengths. Korista
\& Goad (2001) found that 20\%--30\% of the observed effect may be due
to reverberation of the diffuse continuum emission of the broad line
clouds. In the simulations presented here, the SED of the continuum
incident on the broad-line clouds is the same for both high and low
continuum states. How an emission line's responsivity would change with
an {\em incident} SED that changed shape as the UV-optical continuum
level varied will depend on the details of how the flux of photons important
to the creation and destruction of the line varies. We briefly discuss
these effects here.

The shape of the ionizing continuum determines the heating for
given values of gas density, ionizing photon flux, and chemical
abundances. All else being equal, harder incident continua result
in higher electron temperatures, and so major cooling lines (e.g.,
\ion{C}{4} $\lambda$1549) might be expected to increase their energy
output in response. One should also keep in mind that the incident
flux at energies corresponding to the {\em Balmer} continuum is
important to the destruction of hydrogen emission lines for the high
gas densities encountered in the BLR (Shields \& Ferland 1993). Also,
harder continua during brighter UV-opt continuum states might also mean
comparatively more photons above 1~ryd and/or 4~ryd, favoring
the production of hydrogen and helium recombination lines. Kaspi \&
Netzer (1999) considered an SED whose break near 4~ryd moves
to higher energies for brighter UV continuum states (their Figure~8d),
with the effect of increasing the 4~ryd/1~ryd photon flux ratio
with increasing $\lambda$1350 continuum luminosity. However, in this
variable SED scheme, the flux above 20~ryd (270~eV) remained
constant. A comparison of their Figs.~7 (fixed SED) and 9 (variable
SED) that show their model predictions of five UV emission lines' light
curves versus observations illustrates some of the aforementioned effects
of a variable continuum SED on line responsivity. While they included
\ion{He}{2}~$\lambda$1640 in their simulations, such was not the case
for the Balmer lines. Their prescription for a variable continuum SED
increased the responsivities of the \ion{He}{2} and Ly$\alpha$ lines,
whereas those of \ion{C}{4}~$\lambda$1549, \ion{C}{3}] $\lambda$1909, and
\ion{Mg}{2} $\lambda$2800 decreased, presumably because the variations
in the continuum beyond $\sim$~150~eV were smaller in their variable
SED scheme than for the constant SED. It is also apparent that the lags
of \ion{He}{2} and \ion{C}{4} increased in the higher continuum states,
because of a shift to larger responsivity-weighted radii.

\subsection{The Importance of the Recombination Lines to the Method of
Quasar Tomography}

The method of ``quasar tomography,'' proposed by Horne et~al.\
(2003), is a unification of reverberation mapping and photoionization
physics. Its goal is to place physical constraints on the recovered
two-dimensional transfer function map $\Psi(\tau, v)$ of the broad
emission lines, as well as to recover the gas distribution described
by a five-dimensional map $f(r, \theta, n_H, N_H, v)$, where $\tau$
is the time delay, $v$ is the observed radial velocity, $r$ is the
distance from the continuum source, and $\theta$ is the angle from
the line of sight. The sensitivity of the emissivities (EW contours in
Figure~1) of the optical recombination lines to the continuum flux, and the
resulting consequences to their responsivities, make these lines important
candidates for inclusion in the quasar tomography. Horne et~al.\ found
that when applied to the mostly collisionally excited UV emission lines,
a certain amount of ambiguity arises in the recovery of the maps due to
the weak dependencies in their EWs (and so responsivities) along lines
of constant ionization parameter ($\propto \Phi_H/n_H$) corresponding
to their maximum emissivities. Their emissivities and responsivities are
largely dependent on $U_H$, with smaller dependencies in gas density due
to de-excitation or thermalization. The optical recombination lines,
along with \ion{Mg}{2} $\lambda$2800, offer additional constraints
in determining the physical conditions within the BLR, in that their
emissivities and responsivities are dependent on the local continuum
flux and so radius for a fixed central continuum luminosity, as well as
changes in this flux as the central continuum source varies.

\section{SUMMARY}

The response of the optical recombination lines to continuum variations,
especially the Balmer lines, depends critically on the time-averaged
photon flux incident on a spatially extended BLR. This strong
dependence exists because of the enhanced excited state populations
(and so optical depths) that occur at higher continuum flux levels,
given the high gas densities found in the BLRs of AGNs. The emissivities,
and hence responsivities, $\eta$, of these lines are generally strongly
anticorrelated with the continuum flux, so $\eta$ changes with distance
within the BLR for a fixed continuum luminosity and with time as the
continuum source varies. Below are the major implications, most of which
provide explanations for observed relationships:

\begin{enumerate}

\item {We predict an intrinsic Baldwin effect due to this non-linear
response, specifically $\eta < 1$, in most emission lines, including
the Balmer lines, even after accounting for geometric dilution
due to reverberation. This is a finding that has only recently been
observationally confirmed. Regarding the optical recombination lines, this
effect should be small for \ion{He}{2} $\lambda$4686 (with responsivity
$\eta \approx 1$), but significant for the Balmer lines. }

\item {We predict that the higher order Balmer lines are more responsive,
$\eta(H\gamma) > \eta(H\beta) > \eta(H\alpha)$, as is generally observed,
and so the broad-line Balmer decrement will steepen in low continuum states,
and flatten in high states. Moreover, \ion{He}{2} $\lambda$4686 should be
substantially more responsive ($0.9 \lesssim \eta \lesssim 1$) than the
Balmer lines, and so large variations in the broad \ion{He}{2}/H$\beta$
integrated flux ratio that are correlated with the continuum state can
be expected, even in the absence of a varying continuum SED. }

\item {A Seyfert~1 galaxy's sub-classification as Seyfert~1.0,
1.5, 1.8, or 1.9 may at least in part be due to the relative flux
of ionizing photons present within their broad-line regions at the
time of observation. Likewise, any given spectrum of a Seyfert Type~1
AGN may pass through some of these sub-classifications as its continuum
luminosity fluctuates from high (Seyfert~1 or 1.5) to low continuum states
(Seyfert~1.8 or 1.9). }

\item {The emission-line responsivity, $\eta$, places an additional
important constraint on the run of physical conditions within the BLR
(e.g., gas density, column density, covering fraction) as a function
of radius over and above that of the line emissivity and lag. This
is because for the optical recombination lines, the mean responsivity
declines with increasing incident photon flux. Thus, for a particular
continuum luminosity state, a smaller emissivity-weighted radius is
accompanied by a lower effective responsivity. }

\item {Greater responsivity in the optical recombination lines is expected
during low continuum states, and lower responsivity is expected during
high continuum states. The characteristic size of the BLR as measured by
the emission-line lag also will track the continuum luminosity level,
a phenomenon known as ``breathing.'' Responsivities that are functions
of the continuum level and hence time result in non-stationary line
transfer functions. In this case the time dependence is due to changes in
the time-averaged photon flux incident on a spatially extended BLR. The
stronger response due to an elevated responsivity at low continuum levels
will be enhanced by the greater coherence of an effectively smaller BLR
due to breathing. This time-dependence in the line transfer functions
should also manifest itself in changes in the line profile widths, with
the broad-line component becoming narrower/peakier in high continuum
states. }

\item {Because the emissivities and responsivities of the optical
recombination lines are anticorrelated with the incident continuum flux,
observations that H$\alpha$/H$\beta$ and H$\beta$/\ion{He}{1} flux ratios
and the Balmer line responsivity decrease from the core to the wings
indicate that the BLR velocity field diminishes with increasing distance
from the central continuum source. These findings are consistent with
recent reverberation studies that find a strong correlation between the
emission-line lag and rms profile width for multiple lines in individual
AGNs, which implies that the gas kinematics are dominated by the central
massive object. }

\item {Larger variations in the Balmer decrement with continuum state will
be found in the broad-line cores rather than in the wings. Observations
of such would also indicate that the BLR velocity field diminishes with
increasing distance from the central continuum source.}

\item {Differences in broad-line widths (whether rms or mean profile) and
lags among the Balmer and helium recombination emission lines are in
large part due to a radial (continuum flux-dependent) stratification of
optical depth effects in these lines. }

\item {Much of the observed behavior of the optical recombination lines
(e.g., the weak response of the Balmer line wings) previously attributed
to a substantial contribution from matter-bounded clouds at small BLR
radii (i.e., the so-called VBLR) may be explained by the photon flux
dependencies of these lines' emissivities and responsivities within
ionization-bounded clouds. }

\end{enumerate}

These findings should be general, largely independent of any
specifically adopted model of the BLR. However, the details of the
optical recombination lines' responsivities will be sensitive to and so
diagnostic of specific physical conditions within the BLRs of AGNs. In
addition to the recombination lines investigated here, we suggest that
studies of the IR Paschen and Bracket lines should prove fruitful in
further constraining the BLR, particularly in its outer regions.

%
\bigskip
\acknowledgements

This paper benefited from the careful and critical review of an
anonymous referee. We are grateful to Gary Ferland for maintaining
his freely distributed code, Cloudy. We thank the University of St.\
Andrews and Keith Horne for their hospitality and support through a
PPARC visitors grant. We also thank Keith for discussions of work in
progress. M.R.G. would like to thank the hospitality and support of
Western Michigan University during the completion of this work.

%
%

\clearpage
\begin{appendix}
{Appendix: SENSITIVITY OF EMISSION LINE RESPONSIVITY TO MODEL ASSUMPTIONS} 

We briefly discuss the point that most of the major trends in line
responsivity reported here should not depend critically on the particular
choice of model for the broad emission-line region. As discussed in
$\S$~2, most of these effects can be understood from Figure~1, which shows
quite generally the dependence of an emission line's emissivity on cloud
density and incident ionizing photon flux, irrespective of how these
clouds (or emitting entities) are distributed or other details.

\section{Cloud Column Densities} 

Korista et~al.\ (1997) discusses and illustrates the effects of
cloud column density on the line emissivities, and thus EW contours in
Figure~1. The present simulations assume a constant cloud column density
of $10^{23}$~cm$^{-2}$. In general, the computed integrated AGN line
spectrum is relatively insensitive to this parameter for column densities
exceeding $10^{22}$~cm$^{-2}$ (see also Goad \& Koratkar 1998), except
for the high ionization lines of \ion{O}{6} $\lambda$1034 and \ion{Ne}{8}
$\lambda$774, which require column densities $\gtrsim 10^{22.5}$~cm$^{-2}$
and $\gtrsim 10^{23}$~cm$^{-2}$ to fully form within the cloud (Korista
et~al.\ 1997; Hamann et~al.\ 1998). To the extent that cloud column
density has a significant effect on the luminosities or responsivities
of the emission lines plotted in Figure~1, the clouds must lie in the
vicinity of the central diagonal boundary representing the steep downward
gradient in EW (i.e., near $\log \Phi_H - \log n_H \sim 10.25$). These
would be the clouds affected by the passage of an ionization front. If the
preponderance of clouds have a smaller column density than those computed
here, this {\em diagonal} boundary would shift toward the lower right
corners of the density-flux planes in Figure~1. For higher column density
clouds, this boundary would shift toward the upper left. The emissivities
of clouds lying far from this boundary in the density-flux plane would
be essentially unaffected. This is clearly illustrated in Figure~3 of
Korista et~al.\ (1997). From Figure~4 and Eq.~2 in the present paper, it
can therefore be seen that changing the cloud column density by an order
of magnitude in either direction from the adopted value will not change
the overall trends of the EW and responsivity of the ionization-bounded
clouds being proportional to the incident continuum flux for the four
emission lines studied here. Because of their inefficient natures,
matter-bounded clouds cannot contribute substantially to the hydrogen
recombination line luminosity (or responsivity), unless they have much
larger covering fractions than the optically thick clouds at the same
radius. Even then, their importance is likely to be minor, except perhaps
in the extreme wings given a velocity field that declines with distance
from the central source.


\section{Incident Continuum SED, Gas Abundances} 
Different assumptions about the incident continuum SED or gas abundances
do not change the line EW distributions in Figure~1 very much (see
Korista et~al.\ 1997), and so would have little effect on these lines'
responsivities.

\section{Optical Depth and Radiative Transfer Effects} 
If the optical depths in the Balmer lines were significantly reduced
by the presence of large extra-thermal velocities that might occur in
an accretion disk wind (e.g., Murray \& Chiang 1997), the result would
be the EW contours in Figure~1 peaking up at higher incident fluxes and
densities and the Balmer line-emitting region being more strongly
influenced by gas with high responsivity ($\eta \approx 1 - 2$). Since
the Balmer lines have reported responsivities less than 1 even after
correcting for geometric dilution, this would constrain models with
large extra-thermal velocities and/or Balmer emission from large outer
radii. Other consequences of significant extra-thermal velocities
might be smaller predicted lags and broader predicted line profiles
(for a fixed central mass and incident continuum luminosity). If in
general the radiative transfer of the Balmer lines is currently less
than adequately handled (as may be the case; see Netzer et~al.\ 1995),
the optical depth of H$\alpha$ should nevertheless remain greater than
that of H$\beta$, and those of the helium lines should remain smaller
than the Balmer lines. Thus, the relative senses of the EW contours of
the Balmer lines in Figure~1 should be preserved. The sense of the Balmer
decrement variability as a function of continuum state should also
be preserved, as should the H$\alpha$/H$\beta$ intensity ratio across
the line profile: {\em provided that the bulk of the velocity field
diminishes with increasing distance from the central source}.

\section{Possible Alternate Trends in Lag and Line Response} 
Most of the extra responsivity in H$\beta$ over H$\alpha$ in Figure~7
lies within the line core whose contributions are dominated by gas
lying (and responding) at larger distances. As a consequence, the
difference in the measured lags of these two lines may not be as
large as their EW distributions in Figure~1 might indicate. That is,
their responsivity-weighted radii may be closer in size than their
emissivity-weighted radii (Goad et~al.\ 1993). In fact, depending on
the actual emissivity and responsivity distributions of the two lines,
it is possible for H$\beta$ to lag slightly behind H$\alpha$. In other
words the total light profile may be broader in H$\beta$ than H$\alpha$
and the other way around in the rms variation profile, with H$\alpha$
having a slightly smaller lag. The small differences in the Balmer
lines' beaming functions (O'Brien et~al.\ 1994) may also play a role
in altering the relative responsivity-weighted radii, and thus relative
lags of these emission lines.

\end{appendix}
%
%

%
\clearpage
\pagebreak

\begin{figure}
\figurenum{1}
\plotfiddle{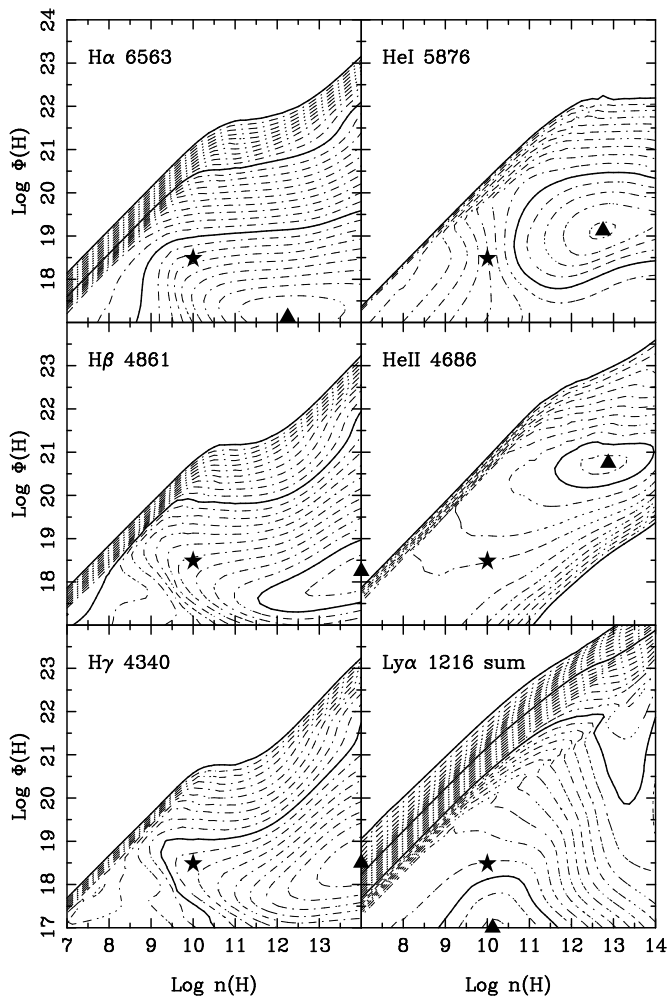}{1.5in}{0}{250}{375}{100}{440}
\caption{Contours of $\log W_{\lambda}$~(EW) for six emission lines,
{\em referenced to the incident continuum at 1215}~\AA\/, are shown as
a function of the hydrogen number density and flux of hydrogen-ionizing
photons for full source coverage at every position. The total hydrogen
column density is $10^{23}~\rm{cm^{-2}}$. The EW is in direct proportion
to the continuum reprocessing efficiency. The smallest decade contour
(generally running diagonally through the centers of the diagrams)
corresponds to 1~\AA\/, each solid line is 1 decade, and dash-dotted
lines represent 0.1 decade steps. The contours decrease monotonically from
the peak ({\em triangle}) to the 1~\AA\/ contour, and the blank regions
in the upper left portion of the panels have EWs less than 1~\AA\/. The
star in each panel is a reference point marking the old ``standard
BLR'' parameters (Davidson \& Netzer 1979).}
\end{figure}

\begin{figure}
\figurenum{2}
\plotfiddle{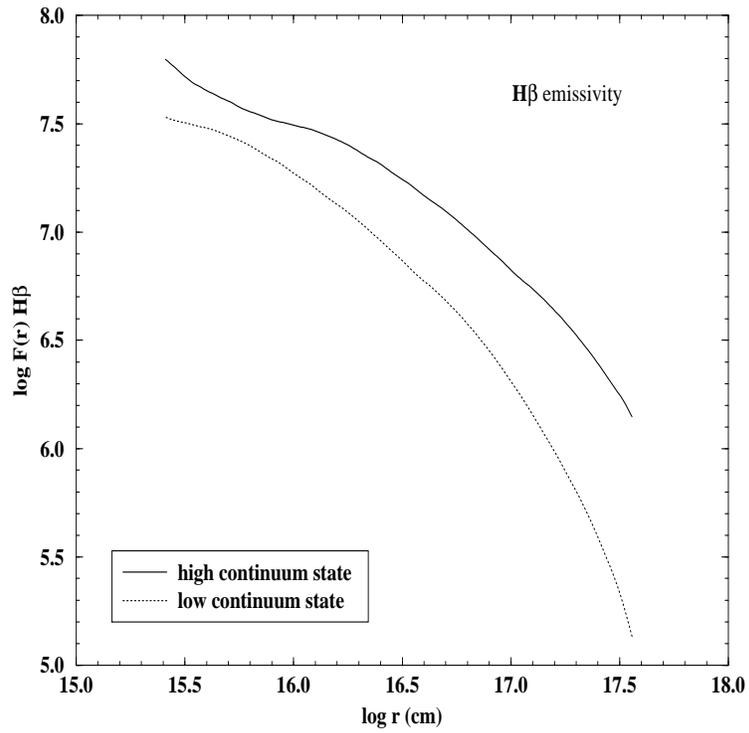}{3.25in}{270}{280}{280}{100}{440}
\caption{Effective logarithmic H$\beta$ surface flux ($\rm{ergs~s^{-1}
cm^{-2}}$) versus the logarithm of the distance from the ionizing continuum
source for the two continuum states. The H$\beta$ emissivities of clouds
with a range of gas densities have been summed at each radius.}
\end{figure}

\begin{figure}
\figurenum{3}
\plotfiddle{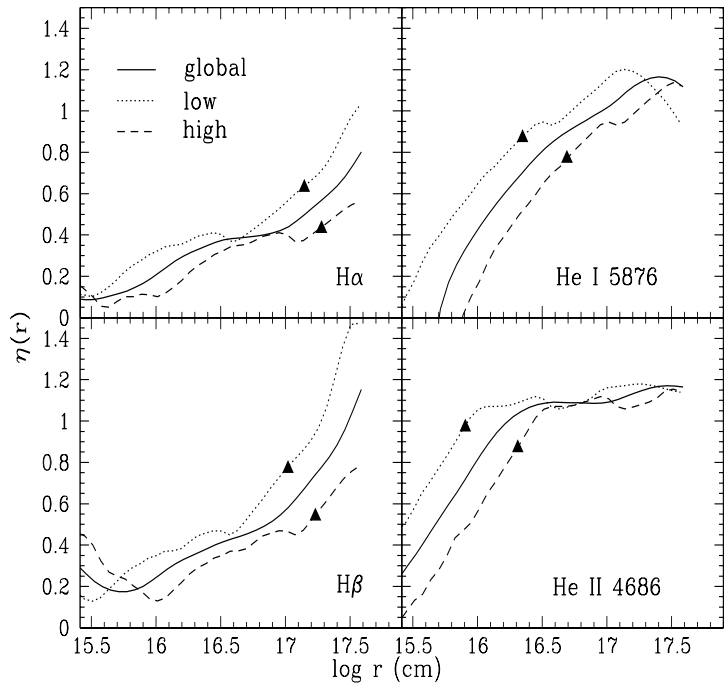}{3.25in}{270}{280}{290}{100}{440}
\caption{Radial responsivities of the four recombination lines under
study. The solid line represents the responsivity for the factor of 8.2
change in continuum flux. The dotted line represents the same for 10\%
variations about the low continuum state, and the dashed line for 10\%
variations about the high continuum state. The responsivity-weighted
radius is marked with a triangle for each of the latter two cases.}
\end{figure}

\begin{figure}
\figurenum{4}
\caption{Same as Fig.~1, but with a superposed gray scale in responsivity.
White represents $\eta < 0$. The lightest shade of gray represents $0 \leq
\eta < 0.5$, medium gray $0.5 \leq \eta < 1.0$, medium dark gray $1.0 \leq
\eta < 1.5$, dark gray $1.5 \leq \eta < 2.0$ (appearing in \ion{He}{2}
$\lambda$4686, H$\beta$, H$\gamma$ only), and black $ \eta \geq 2$
(appearing only in the panel for \ion{He}{2} $\lambda$4686, which shows
all of the shades). These responsivities were computed for 0.45 decades
(factor of 2.8) variations in the continuum luminosity about the adopted
mean continuum state.}
\end{figure}

\begin{figure}
\figurenum{5}
\plotfiddle{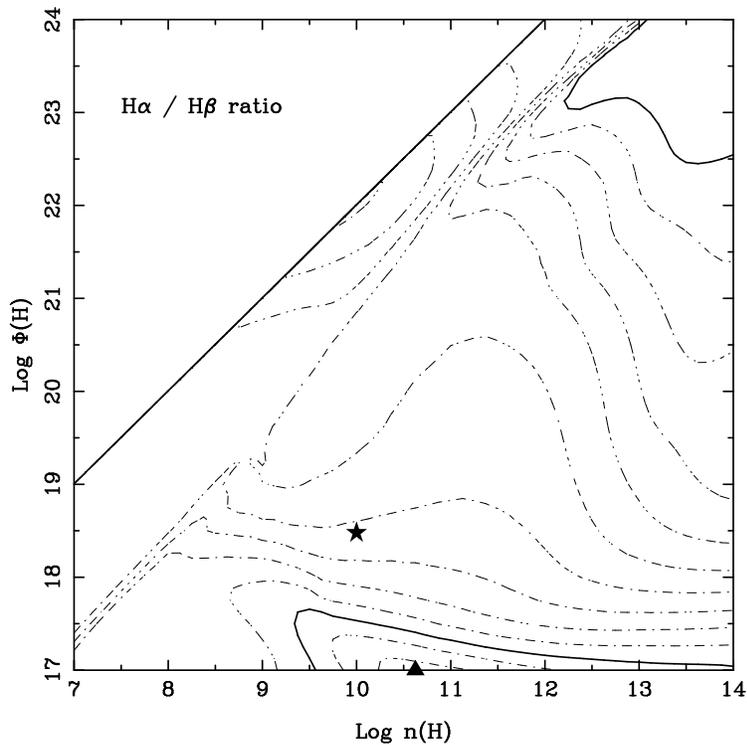}{3.25in}{0}{280}{280}{100}{440}
\caption{Logarithmic flux ratios of H$\alpha$/H$\beta$ in the density-flux
plane. The solid contour near the bottom of the diagram represents a
flux ratio of 10 (1.0 in the log), the dash-dotted contours are 0.1
decade intervals, and the solid contour in the upper right corner
represents a flux ratio of 1.0 (0.0 in the log). The triangle lies at
the peak ratio of $\approx 17$. The star is a reference point marking
the old ``standard BLR'' parameters of Davidson \& Netzer (1979), and
has a value of $\approx 4.2$. The contours end abruptly at the central
diagonal solid line, where the equivalent widths of the two lines become
very small due to the extremely high ionization state of the gas.}

\end{figure}

\begin{figure}
\figurenum{6}
\plotfiddle{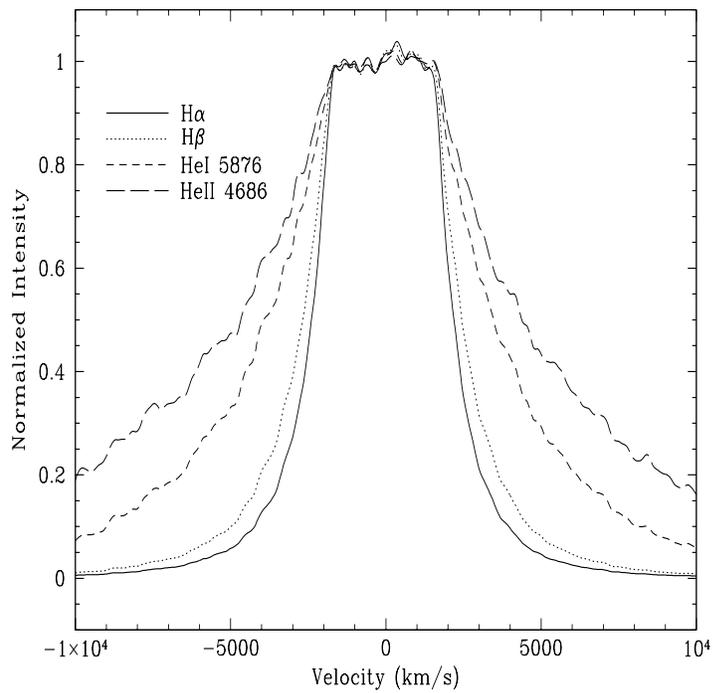}{3.25in}{270}{280}{280}{100}{440}
\caption{Model emission-line profiles for the four recombination lines
under study, scaled to their peak intensities to ease comparison. See
$\S$~3.5 for details.}
\end{figure}

\begin{figure}
\figurenum{7}
\plotfiddle{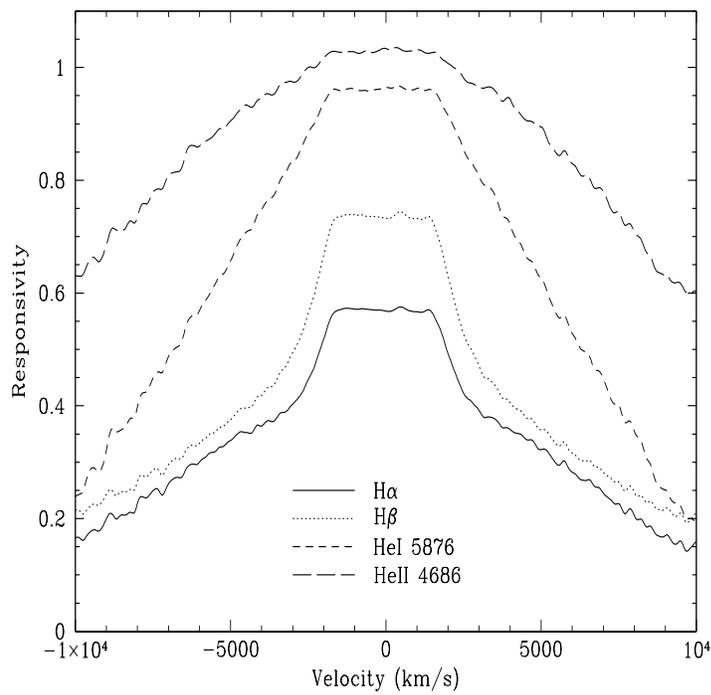}{3.25in}{270}{280}{280}{100}{440}
\caption{Model emission-line responsivities presented as functions of
radial velocity for the four recombination lines under study ($\S$~3.5).}
\end{figure}

\begin{figure}
\figurenum{8}
\plotfiddle{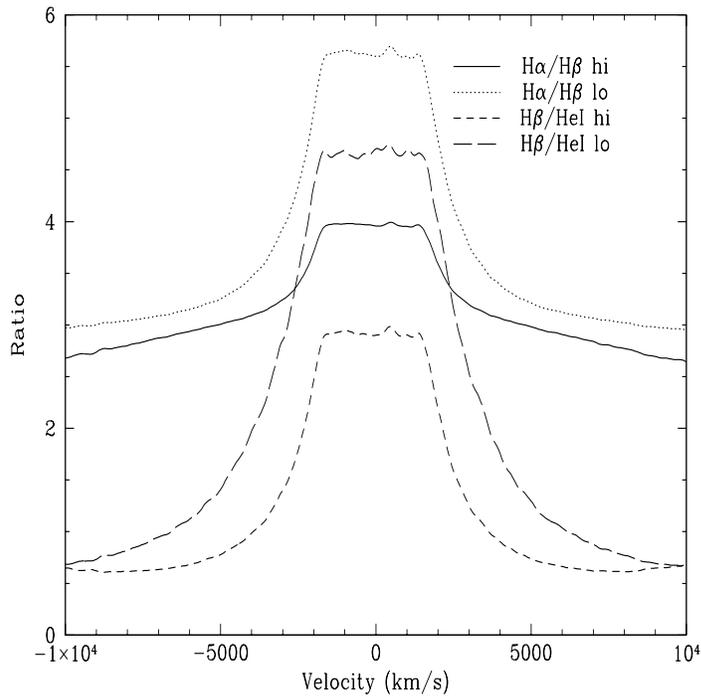}{3.25in}{270}{280}{280}{100}{440}
\caption{H$\alpha$/H$\beta$ and H$\beta$/\ion{He}{1} $\lambda$5876 flux
ratios presented as functions of radial velocity in the low and high
continuum states ($\S$~3.6). In both cases, the smaller ratios are found
in the high continuum states because of the stronger response of the line
in the ratio's denominator.}
\end{figure}


\clearpage

\begin{table}
\caption{Model Effective Line Responsivity \& Profile Variability }
\begin{tabular}{lccccc}
\tableline
\multicolumn{1}{c}{Emission Line} &
\multicolumn{1}{c}{$\eta\/_{eff}$(global)} &
\multicolumn{1}{c}{$\eta\/_{eff}$(hi)} &
\multicolumn{1}{c}{$\eta\/_{eff}$(lo)} &
\multicolumn{1}{c}{FWHM(hi) (km/s)} &
\multicolumn{1}{c}{FWHM(lo) (km/s)} \\
\tableline
H$\alpha$ $\lambda$6563 & 0.52 & 0.43 & 0.63 & 4200 & 4600\\
H$\beta$ $\lambda$4861 & 0.64 & 0.54 & 0.77 & 4510 & 5230\\
H$\gamma$ $\lambda$4340 & 0.68 & 0.58 & 0.79 & 4570 & 5250\\
\ion{He}{1} $\lambda$5876 & 0.83 & 0.77 & 0.87 & 5780 & 7350\\
\ion{He}{2} $\lambda$4686 & 0.92 & 0.87 & 0.97 & 7590 & 9480\\
Ly$\alpha$ $\lambda$1216 & 0.74 & 0.71 & 0.77 & 5180 & 5280\\
\tableline
\tableline
\end{tabular}
\end{table}


\end{document}